\documentclass[12pt]{article}

\usepackage{scicite}
\usepackage{graphicx}
\usepackage{times}
\usepackage{xcolor}
\usepackage{hyperref}
\usepackage{floatrow}
\newfloatcommand{capbtabbox}{table}[][\FBwidth] 

\newenvironment{sciabstract}{%
\begin{quote} \bf}
{\end{quote}}

\title{It Takes So Little to Change So Much: Investigating the Robustness of a Danish Voting Advice Algorithm}

\author{Giovanni Astante$^{1,2}$, Roberta Sinatra$^{1,3,4}$, and Vedran Sekara$^{1,4\ast}$\\
\normalsize{$^{1}$Networks, Data and Society (NERDS), IT University of Copenhagen, Denmark}\\
\normalsize{$^{2}$Department of Informatics, University of Zurich, 
Switzerland}\\
\normalsize{$^{3}$Center for Social Data Science (SODAS), University of Copenhagen, Denmark}\\
\normalsize{$^{4}$Pioneer Centre for Artificial Intelligence (P1), Copenhagen, Denmark}\\
\normalsize{$^\ast$To whom correspondence should be addressed e-mail: vsek@itu.dk}
}

\date{}

\begin{document} 


\baselineskip20pt


\maketitle


\begin{sciabstract}
Voting Advice Applications (VAA) are tools designed to help voters compare political candidates on policy preferences prior to elections.
VAAs are popular tools in European countries and in other countries with multi-party democratic systems. 
Through a freedom of information request we got access to the inner workings of a popular Danish VAA called the `\textit{Kandidattest}' which is implemented by a major Danish news outlet and has been used for general, municipal, and European elections. Users and politicians from every political party answer the same online questionnaire and get matched based on the agreement percentage stemming from their answers.
VAAs play a significant role in elections with 45\% of surveyed voters reporting they followed their recommendations in the past Danish general election. However, the inner workings of VAAs have not been thoroughly evaluated until now.
We find that the algorithm is not robust enough for users to trust the agreement percentages in the output, as small changes to the algorithm can lead to different results, potentially affecting election outcomes.
We conduct an algorithmic audit of the \textit{Kandidattest}'s robustness, using simulated responses to investigate the tool's brittleness, with respect to minor adjustments of the algorithm's weight, and changes in the number of questions in the questionnaire.
\end{sciabstract}


\section{Introduction}
Algorithms profoundly impact our daily lives~\cite{willson2019algorithms}.
They shape our social networks, mediate our romantic lives, and are even used to help us make political decisions.
Voting Advice Applications (VAAs) are online tools used in countries with multi-party systems~\cite{cedroni2010voting} to help voters compare political candidates across their policy preferences and identify politicians, or political parties, they could potentially vote for.
VAAs are services typically hosted on web pages that match users and politicians, according to similarities found between their answers to the same administered questions. They are typically hosted by media companies or other public sector oriented organizations, and are widely used.
For example, in the Netherlands the most popular VAA was used 6.8 million times within an electorate of 12.9 million eligible voters during the 2017 national elections~\cite{munzert2021meta}. The popularity of VAAs, however, has
been found to vary across democracies. From roughly 3\% of the electorate in the 2009 elections to the European Parliament using them, between 10\% to 20\% in recent national elections in Canada, Finland, Germany, Greece, New Zealand, and Switzerland, to one-third of eligible voters in recent Dutch and Danish national elections~\cite{germann2019getting}. Although a recent meta-analysis has pointed out that measurement issues potentially affected some of the above-mentioned results~\cite{munzert2021meta}, it also confirmed that VAAs usage has positive effects on voting turnout, vote choice, and that VAAs can increase the likelihood of voting~\cite{munzert2021meta}. In addition, VAAs are believed to level the playing field between politicians, as they solely rank candidates according to similarity scores, ignoring factors such as popularity of candidates, length of political tenure, and media coverage.
In Denmark, VAA usage is quite common~\cite{KristeligtDagblad, Epinion, Radio4}, with 62\% of eligible voters using VAAs one month prior to the 2022 general elections and 45\% reporting they followed the recommendation of the VAA when voting~\cite{DanishNationalElectionStudy}.

While directly assessing the impact of these technologies on public political behavior is not straightforward, it is important to investigate the potential pitfalls of using an automated system in such a high-risk setting.
Bearing in mind that there is no such thing as a `raw' or neutral piece of technology in terms of it being objective, as every piece of technology is intertwined with various political, technical, cultural, and social aspects~\cite{willson2019algorithms}.
Even the concepts of fairness and justice, when applied to technology, must be anchored to the social context in which any technology is applied~\cite{selbst2019fairness}. Nonetheless, in the context of Denmark past work has demonstrated that VAAs have substantial effects on voter choices~\cite{tromborg2023candidates}. The study found that many VAA users switched to the recommended party, with undecided voters and voters with low levels of political interest being much more likely to follow VAA advice.

It is against this backdrop that we situate our study. 
Although VAAs have been studied through the lens of how people use them, little is known about their robustness: How robust are the recommendations of these systems? Will small changes to their inner workings result in fundamentally different candidates being recommended?
We focus on a Danish VAA as Denmark is one of the most digitized societies in the world~\cite{digitalization}, with high levels of trust in institution and digital solutions, and VAAs being frequently used by voters (62\% of eligible voters used VAAs in the 2022 elections~\cite{DanishNationalElectionStudy}).

Here we analyze a VAA called the `\textit{Kandidattest}' (in english "The Candidate Test") used by a major Danish newspaper and media house (\textit{Politiken}). 
Through a freedom of information request we were granted access to the algorithm they adopted to match users and political candidates for several past elections (general, municipal, and European elections). We obtained source code and documentation from the media house which enabled us to replicate the VAA and study its robustness. The system functions by a user filling out a questionnaire, which is compared to a database of answers from political candidates and the candidate (or top-$k$ candidates, depending on the VAA) with the highest agreement percentage are recommended. Our work aims to audit the robustness of this VAA.
We approach this task from two perspectives. First, we investigate how minor modifications to the algorithm’s weights can change the outcome of which political candidates are recommended. Second, we study how dropping questions, i.e. calculating match similarity based on fewer questions, affects how candidates are recommended. Our investigation focuses both on how outcomes are affected at the level of individual political candidates, and at the level of political parties. Our results demonstrate that minor tweaks to the algorithm can produce entirely different results, potentially affecting the outcomes of political elections. Based on our findings we provide actionable recommendations to VAA developers, and policy makers, on how to monitor and improve the robustness of these systems.

\section{Overview of the \textit{Kandidattest} and its possible outcomes}
The \textit{Kandidattest} was developed between 2012 and 2014 as a joint project by three large Danish media companies (\textit{Politiken}, \textit{Jyllands-Posten}, and \textit{Ekstra Bladet}).
According to the documentation received in our freedom of information request, all three media companies had different versions of the VAA running on their websites during multiple elections.
Here we focus on the version hosted on \textit{Politiken}'s website. We got access to the algorithm used during the 2022 general elections. The algorithm was also used for the 2026 general elections and 2025 municipal elections but with different questions in the questionnaire, as policy issues change over time and differ between municipal and general elections.

\subsection{System overview}
The \textit{Kandidattest} consists of 20 questions selected by journalists from the media house. The questions are tailored to each election and designed to establish the most complete possible picture of a person's political preferences. The questions are also shared with the various political parties, which provide feedback to ensure all relevant policy questions are sufficiently covered in the VAA~\cite{altinget}. Questions can probe multiple political topics, such as labor market, economy, climate change, social welfare, and education, to name a few.
The selected 20 questions are then answered by a large majority of candidates running for election. Every VAA user (or voter) also answers the same questions. 
Users provide answers on an ordered scale ranging from 1 to 5 corresponding to a disagree-neutral-agree range (Likert scale), with 1 indicating the choice `\textit{completely disagree}', and 5 denoting `\textit{completely agree}'.
In addition, users can assign an importance value to each question, indicating how important it is (or how important the topic is). 
Users can choose among three options: \textit{Important}, \textit{Neutral}, or \textit{Not Important} for each question. The importance of the questions is then used to weight answers when calculating similarity scores.

To match users to political candidates, the system computes a distance score for each question between the user's answer and each candidate's answer. Here we denote $d_i = |u_i - p_i|$ as the distance between a user's answer $u_i$, and a political candidate's answer $p_i$ to question $i$. Distances can range from 0 (both the user and the political candidate have answered the same) to 4 (they have provided opposite answers).
To aggregate the total distance between users and political candidates, the algorithm's developers have chosen to use a weighted table lookup.
The weights ($w$) are listed in Table~\ref{tab:weighted_scores}, and are split up according to distances $d_i$, and the importance of questions $I_i$.
The total distance between a user and a political candidate is calculated as the total weighted distance across all $m=20$ questions:
\begin{equation}
\textrm{total weighted distance} = \sum_{i=1}^{m} w(d_i, I_i) \label{eq:distance}
\end{equation}

In our case, the total weighted distance ranges from 0 to 720. 
To convert these values to a scale understandable to laypeople, the developers of the algorithm normalized them to the maximum possible total weighted distance. For the weights in Table~\ref{tab:weighted_scores} the maximum weighted distance is $720$, determined by multiplying the highest weight, $w(\textrm{d = 4},\textrm{I = \textit{Important}}) = 36$ by $m=20$ questions. This is a special variant of min-max scaling, although the developers never mention it in the provided documentation. The disagreement percentage is calculated as:
\begin{equation}
    \textrm{disagreement percentage}  = 100 \times \lceil \frac{\textrm{total weighted distance}}{\textrm{max weighted distance}} \label{eq:disagreement}
\end{equation}
Here, $\lceil$ denotes a ceil operator, which rounds the result to the closest integer greater than or equal to the number.
For example, a value of 15.1 is rounded up to 16.
Disagreement is then transformed to agreement by subtracting it from 100:
\begin{equation}
    \textrm{agreement percentage} = 100 - \textrm{disagreement percentage} \label{eq:agreement}
\end{equation}
The agreement scale ranges from 100 (complete agreement) to 0 (complete disagreement). These values are reported as percentages when presenting the results back to users. For example, a value of $80$ is presented as $80\%$.

\begin{figure}[b]
    \centering
    \includegraphics[width=\textwidth]{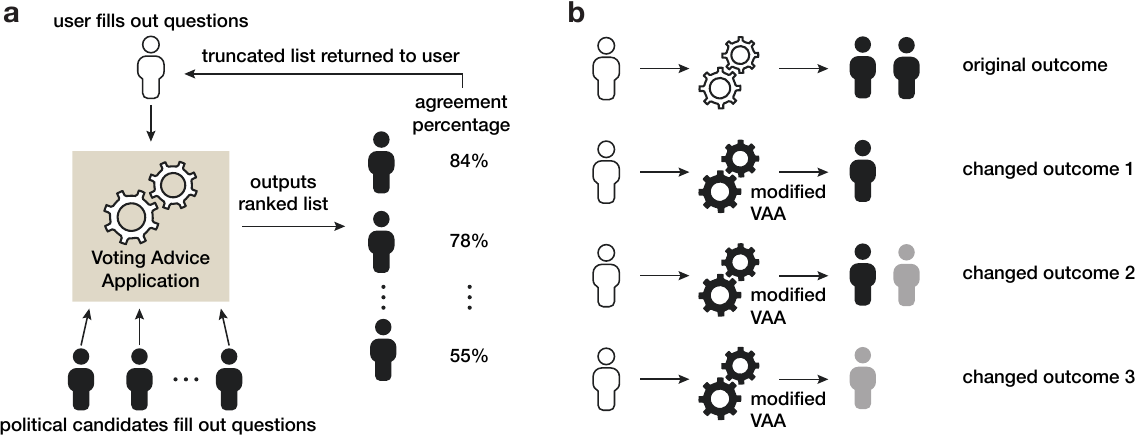}
    \caption{Characteristics of the \textit{Kandidattest}, and possible different outcomes. \textbf{a}, Visual overview of the matching algorithm. A user fills out $m$ questions on a webpage and the matching algorithm compares their answers to answers by political candidates. A ranked list is returned to the web server which presents a truncated list of these to the user. \textbf{b}, Visual explanation of possible different outcomes. When modifying the algorithm we compare outcomes to the original version of the \textit{Kandidattest}. If any of the three possible outcomes occur we say a marginal modification in the algorithm has changed the outcome of the model.}
    \label{fig:fig1}
\end{figure}

Results are calculated for all user $\times$ political candidate pairs, and are presented to users on a web page that shows the political candidates with the highest agreement percentages to the user, together with their political party affiliations.
A visual representation of the system is shown in Fig.~\ref{fig:fig1}a.
The final results are shown to users, with a major focus on the political candidate with the highest agreement score: their picture and the agreement percentage occupy a large portion of the visual space. In the case of ties, when identical agreement percentages are shared by two candidates, multiple candidates are shown.
Names, pictures, and agreement percentages associated with other runner-up candidates are shown as well, but with smaller dimensions. 

\begin{table}[]
\resizebox{\textwidth}{!}{
\begin{tabular}{c|c|c|c|c|c}
\textbf{Importance}                 & \textbf{Score Distance 0} & \textbf{Score Distance 1} & \textbf{Score Distance 2} & \textbf{Score Distance 3} & \textbf{Score Distance 4} \\ \hline\hline
\multicolumn{1}{l|}{Not Important} & 12                        & 15                        & 18                        & 21                        & 24                        \\ \hline
Neutral                             & 6                         & 12                        & 18                        & 24                        & 30                        \\ \hline
Important                           & 0                         & 9                         & 18                        & 27                        & 36                        \\ 
\end{tabular}
}
\caption{The weights of the matching algorithm ($w(d_i,I_i)$), used in the `\textit{Kandidattest}' as provided by a freedom of information request from the news organization \textit{Politiken}. Each weight is used in the calculation of Eq.~\ref{eq:distance}. Why weights are multiples of 3 is unclear, and is never mentioned in the documentation.}
\label{tab:weighted_scores}
\end{table}

\subsection{Defining matching outcomes}
The robustness of a VAA can be quantified in different ways. Previous studies have investigated the impact of how questions are phrased and which policy questions are selected for the questionnaire~\cite{walgrave2008vote}; which computational methods are used to calculate matches~\cite{louwerse2014design}; and how results are presented and graphically visualized~\cite{bruinsma2020evaluating}. We test how marginal changes to the algorithm can potentially affect matching. We focus on two types of change: i) how marginal changes in the algorithm's weights affect matching, and ii) how removal of $n$ of the original $m=20$ questions affects matching. To quantify the effects, we compare recommendations from a modified algorithm to recommendations from the original algorithm. 
We consider the outcome to change when the top-ranked candidate is different from the original matching (Fig.~\ref{fig:fig1}b). 
In the case of ties, where two or more candidates share the same maximum agreement percentage, we incorporate all of them in the analysis.

Fig.~\ref{fig:fig1}b illustrates the possible different outcomes that can arise when we modify the matching algorithm.
In the example, the original outcome is a tie between two candidates. If any of the different possibilities in Fig~\ref{fig:fig1}b occur, we say the marginal modification in the algorithm has significantly affected the outcome of the VAA.
This includes: 1) if one of the candidates in the tie has been dropped by modifying the algorithm. This occurs when their agreement percentage is lower than the other candidate's (even by a difference of 1 percentage point).
2) If a new candidate appears in the tie, or 3) a completely new candidate appears as the top-ranked candidate.
Overall, we only focus on how marginal modifications of the algorithm affect how candidates are ranked; meaning we disregard the agreement percentages of candidates.

So far, we have discussed how the ranking of candidates is affected by modifying the VAA. 
However, candidates can belong to different political parties, which is important to take into account. For instance, the modified algorithm might return a different top-ranked candidate, compared to the original algorithm, but if both candidates are from the same political party then it might not make sense to deem this change as significantly different. 
We can narrow this requirement and only say that a significantly different outcome occurs when candidates from a different political party are ranked on top.
For this reason, we focus on two versions of changed outcomes:

\begin{itemize}
    \item \textit{Changes at candidate level}: we study how modifications to the VAA algorithm, compared to its original version, affect the ranking of candidates. If any of the changes listed in Fig.~\ref{fig:fig1}b occurs we say the outcome has changed, regardless of the candidate party membership. 
    
    \item \textit{Change at political party level}: we study how changes in the algorithm affect how candidates and parties are ranked. A different outcome is defined when the party membership of any candidate in the new output differs from the ones in the original matching. 
\end{itemize}

\section{Quantifying the robustness of the \textit{Kandidattest}}
To estimate robustness, we scraped answers for all political candidates from the \textit{Kandidattest} used for the 2026 and 2022 general elections, and the 2025 municipal elections. The focus of our audit is primarily on the 2026 version, and we also compare these results to the 2022 version. In addition, we perform a `smaller' audit for the 2025 municipal elections, as the VAA was only implemented for the municipality of Copenhagen in that case. 

\begin{figure}[!hb]
    \centering
    \includegraphics[width=0.6\textwidth]{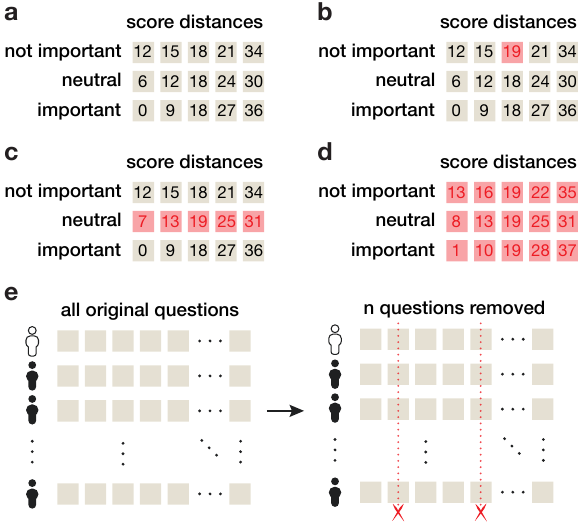}
    \caption{Modifications we perform on the \textit{Kandidattest}. \textbf{a}, Original algorithmic weights split up according to the perceived importance of a question. \textbf{b}, Single modification. Here $w(\textrm{d = 2},\textrm{I = \textit{not important})}$ has been modified by addition of $\delta = 1$. \textbf{c}, Importance weight modification, where all weights for questions marked with a specific importance type are adjusted by $\delta$. Here weighs for questions marked as `neutral' are modified by $\delta=1$. \textbf{d}, Overall modification, where all weights are incremented by addition of $\delta$. \textbf{e}, Removing questions, where $n=2$ random questions have been removed when computing the agreement scores.} 
    \label{fig:fig2}
\end{figure}

Various modifications can be made to the \textit{Kandidattest}, and Fig.~\ref{fig:fig2} illustrates those we made.
Fig.~\ref{fig:fig2}b shows what we call a `single modification' which increments one of the algorithm's weights $w(d_i,I_i)$ by a small amount $\delta$. We only focus on positive values of $\delta$ to ensure weights do not become negative, and only use integer values, specifically $\delta = 1,2,3$, to stay consistent with how weights have been defined by the developers.
Note that when modifying weights, we also dynamically update the maximum weighted distance in Eq.~\ref{eq:disagreement}.

Another modification involved a shift of an entire row of the weights matrix, also called `importance weight modification' (Fig.~\ref{fig:fig2}c). In this case, we add $\delta$ to all score distances for questions a user marked, for example, as `neutral'  ($w(d_i,I=\mathrm{neutral})$).
The last type of adjustment is to increment all weights by a factor of $\delta$, which we call the `overall modification' (Fig.~\ref{fig:fig2}d). This shifts the entire weights matrix by a factor of $\delta$.
Lastly, to investigate how robust the \textit{Kandidattest} is with respect to the number of questions, we also modify the algorithm by dropping $n$ questions when calculating the agreement score. We call this modification `removing questions'.
Fig.~\ref{fig:fig2}e illustrates this change, where $n=2$ questions have been dropped. We drop questions at random for this type of modification.

To calculate agreement scores, we create synthetic users in two ways. First, to explore the full \textit{matching space} of the VAA we create users who answer questions and assign question importance randomly. We call these `\textit{random users}'.
We create 100 batches of 1,000 random users to quantify confidence intervals; that is, we use a total of 100,000 synthetic users to estimate the impact of each modification to the algorithm.
Second, to simulate more realistic responses, we modify the answers from the political candidates by offsetting all their responses with a factor of $+1$ and $-1$. For edge cases where we offset with $+1$ and a candidate has already answered $5$ to a question, we keep the original answer. We do the same for the $-1$ offset and answers of $1$.
This second type of simulation ensures we have real-world answers where possible correlations between responses to different questions are preserved. We denote these as `\textit{real-world users}'.
In this case, since we have no information about how political candidates rank the importance of topics, we study scenarios where all questions have been marked with a single importance category, namely all marked as 'Important', 'Neutral', or 'Not important'.
Below, we first focus our analysis on `random users', followed by a comparison with results from `real-world users'. We end the analysis by quantifying the potential impacts that slight changes to the \textit{Kandidattest} could have had on parliamentary seats.

\subsection{The impact of single modifications}
\begin{figure}[h]
    \centering
     \includegraphics[width=\textwidth]{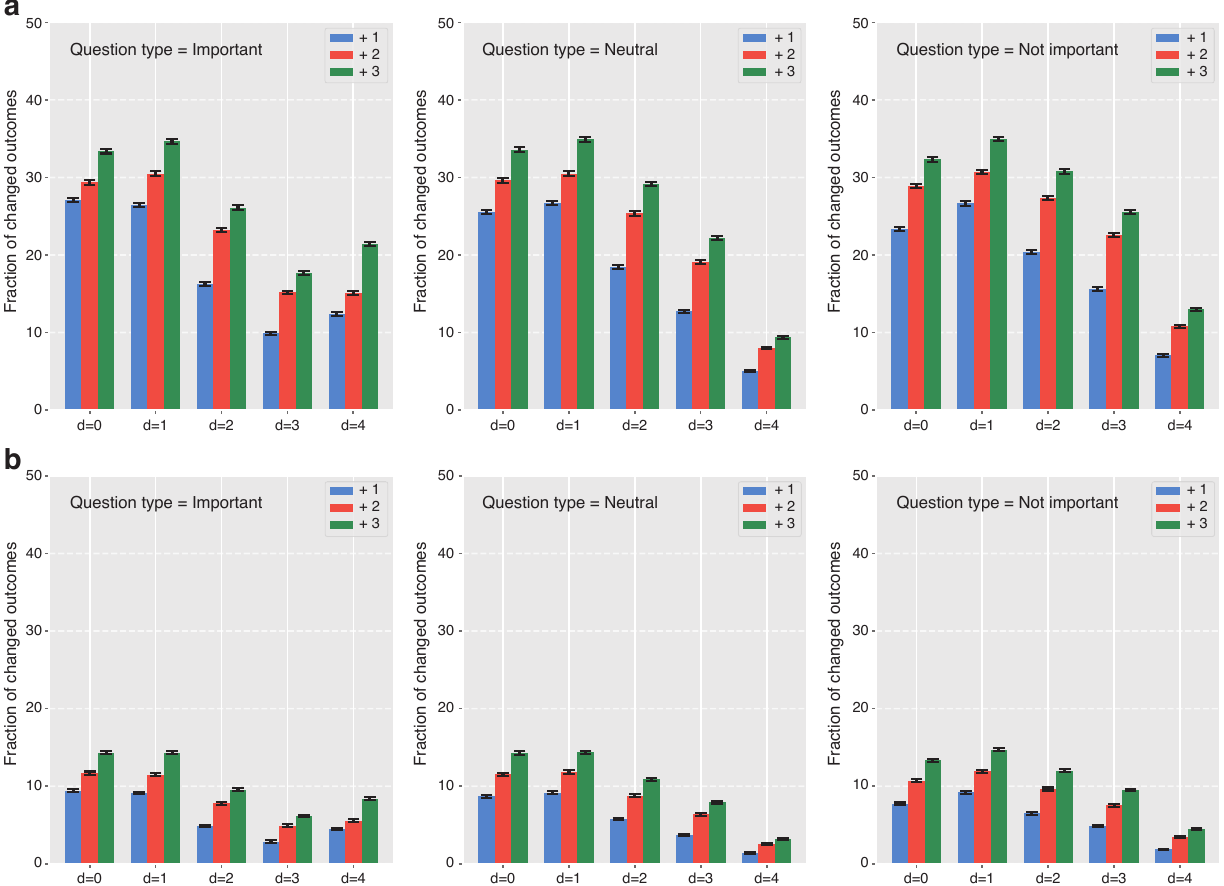}
    \caption{Percentage of changed outcomes as a result of single modifications. Results are for `random users', split up according to importance of questions and different $\delta$ values. Left: questions marked as `important' by users, middle: `neutral', right: `not important'. \textbf{a}, Results for changed outcomes at candidate level. Error bars denote 95\% confidence intervals across different batches. \textbf{b}, Results at political party level. Results for 2022 general elections are nearly identical.}
    \label{fig:fig3}
\end{figure}

Using $\delta = 1, 2, 3$, we edit each individual weight in the algorithm (Fig.~\ref{fig:fig2}b) and measure what happens to the top-ranked outcomes.
Fig.~\ref{fig:fig3}a shows the results for candidate level outcomes for `random users', split up according to question importance, score distance weights, and different values of $\delta$. 
We find that even a marginal modification of $\delta =1$ can have large consequences. 
Independent of whether we change the weights for `important', `neutral', or `not important' questions, a change to zero-distance weights ($w(d=0) + \delta$, the weight added to the total weighted distance when a user's answer aligns with a political candidate) results in respectively 27\%, 25\%, and 23\% of outcomes being different. 
These are significant changes considering we assign question importances at random for `random users', meaning only approximately one third of question will be marked as, for example, `important'.
Unsurprisingly we find that larger $\delta$-values of $2$ and $3$ only increase the amount of outcomes being changed. 
Interestingly, changes to $w(d=1)$ have marginally larger impact on outcomes, but as distances increase the amount of changed outcomes decreases again.
This is a byproduct of the Likert scale employed in the \textit{Kandidattest} and our synthetic user generation strategy, where score-distances of $1$ are more likely to be observed compared to distances of $0$. For example, if a political candidate answers $3$ for a question a random user is more likely to have a score-distance $d=1$ by either answering $2$ or $4$, compared to $d=0$ where the answer has to be exact.
As such, changes to $w(d=1)$ will have slightly larger impacts on changed outcomes.
Nonetheless, Fig.~\ref{fig:fig3}a illustrates that a small changes to any of the weights can greatly influence which political candidate users are matched with.

For the case of political parties, Fig.~\ref{fig:fig3}b shows qualitatively similar patterns, although less extreme. 
It shows that a large fraction of the changed outcomes in Fig.~\ref{fig:fig3}a occur between candidates from the same parties. 
However, modifying $w(d=0)$ with $\delta=1$ still changes approximately 7-9\% of outcomes (depending on question importance). This is disturbingly high.
Near identical patterns were found for the 2022 general version of the \textit{Kandidattest}. For example, $\delta=1$ is associated with 4\% to 28\% changed outcomes when looking at candidates level mismatches - depending on the modified weighted distance and its importance level - and 1\% to 10\% when looking at party level mismatches. 

We repeated the analysis on the \textit{Kandidattest} used during the 2025 municipal elections.
For this version we restrict ourselves to to the capital region of Copenhagen, as the test was only rolled out for that municipality. Moreover, as questions were answered at political party level, and not by individual candidates, we focus only on party level robustness.
The results qualitatively resemble what we have seen in Fig.~\ref{fig:fig3}b, however, the magnitude of changes is higher. A $\delta =1$ modification approximately changes respectively 18\%, 17\% and 16\% of outcomes for important, neutral and not important-type questions.
This is higher than the effect at party level for the 2026 and 2022 general elections because questions are answered at party level, rather than candidate level (16 parties vs. 779 candidates).  
When questions are answered by candidates, a modification to the \textit{Kandidattest} has a possibility of replacing a candidate with a different candidate from the same party. This is not possible when answers are only collected at the party level. In this case the algorithm produces more volatility. 

\subsection{Impact of importance weight modifications}
When modifying all distance weights for all questions of a specific importance-type, that is $w(I) + \delta$ (Fig.~\ref{fig:fig2}c), we find outcomes change qualitatively the same independently of questions being marked as `important', `neutral', or `not important'.
At candidate level, this results in 18\% of outcomes being changed, when $\delta$ = 1.
We find that modifying weights for `neutral' and `non-important' questions results in slightly greater changes than for weights for `important' questions (respectively 18.4\% and 18.2\% compared to 18.1\%); however, as this is within error bars, it is not a meaningful difference. 

Evaluating changed outcomes at party level we find that approximately 7\% of outcomes are changed (specifically 7.3\% for `important', 7.0\% for `neutral', and 7.0\% for `not important' questions).
This illustrates that shifting the weights of any importance-type, while keeping everything else constant, can substantially change which political candidates and parties are recommended. Qualitatively similar patterns are found for the 2022 version of the \textit{Kandidattest}: $\sim$18\% across importance levels for candidate level mismatches and $\sim$7\% for party level mismatches.

\subsection{Impact of overall modifications}
We expect that offsetting all weights with $\delta$\ (Fig.~\ref{fig:fig2}d) should not result in a big change as differences between weights and question importance-types are kept constant. 
Nonetheless, we find that 23\% of outcomes are changed at the candidate level, and 9\% at the party level, with $\delta=1$. 
This is a troubling finding, as this modification preserves the internal logic of the weights while changing all weights equally. It indicates that the algorithm behind the \textit{Kandidattest} has serious robustness issues. 
Again, near-identical patterns are found in the 2022 version of the VAA: $\sim$22\% of outcomes changed for candidate level mismatches and $\sim$9\% for party level mismatches.

\subsection{Quantifying changes in top-$k$ ranked candidates}
\begin{figure}[h]
    \centering
    \includegraphics[width=\textwidth]{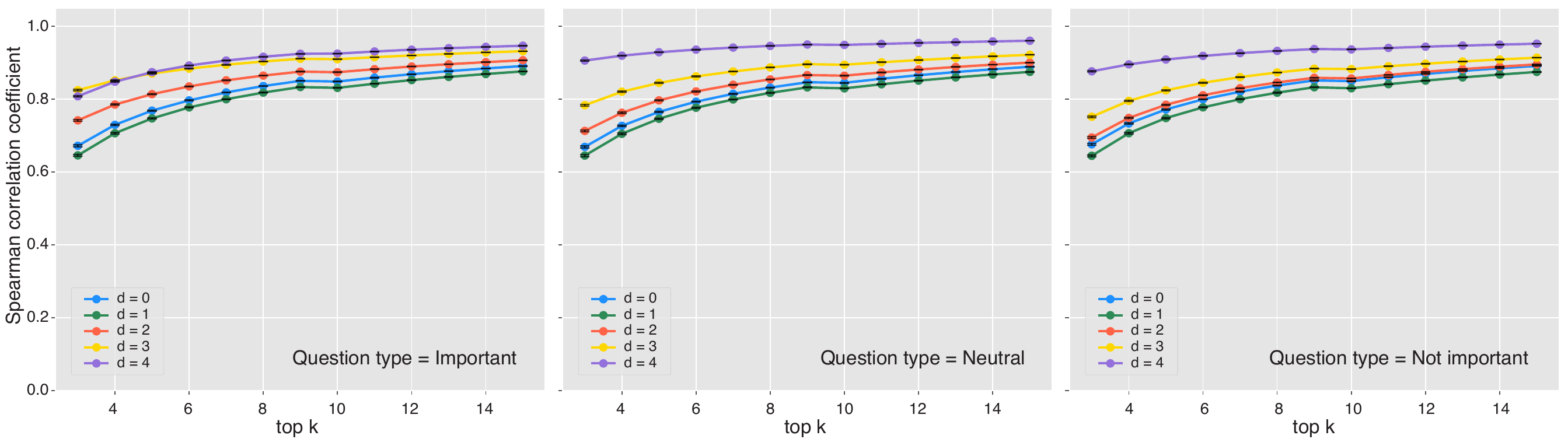}
    \caption{Estimating the robustness of top-$k$ ranked candidates in response to a single ($\delta = 3$) modification of the algorithm's weights. We focus on $k$-values from $3$ to $15$. Results are split up according to question importance-type, and shows changes at candidate level. Error bars denote 95\% confidence intervals across different batches.}
    \label{fig:fig5}
\end{figure}

Until now, we have focused on changed outcomes for the top-1 ranked candidate (or candidates in case of ties). The reason for this is historic, as past versions of the \textit{Kandidattest} used to return only the top match. However, for the 2022 general elections, the VAA returned the top three highest agreement candidates, and this was changed again for the 2025 municipal and 2026 general elections, which provided a larger list of candidates and also a list of parties (where match similarity was computed on the political party level). As such, we also focus on how the list of top-$k$ ranked candidates changes in response to modifications of the algorithm. 
We repeat the `single modification' analysis here from Fig.~\ref{fig:fig3} where we compare outputs from the original algorithm to outputs where a single weight is modified by the addition of $\delta$. 
Fig.~\ref{fig:fig5} shows the results for $\delta=3$, split up according to question importance. We focus on $\delta=3$ to illustrate the worst-case effects.
We use the Spearman correlation (rank correlation) to compare the alignment between ranked lists of candidates. A rank correlation of $1$ indicates perfect correlation, $0$ indicates no correlation, and $-1$ indicates anti-correlation.  

For $k=3$, there are large differences as the correlation coefficient is only approximately $0.65$ when weights $w(d=0)$ and $w(d=1)$ are modified (independent of question importance type). The same percentage is found with the 2022 version of the VAA.
This illustrates a large mismatch between the recommendations of the modified and original algorithms.
Again, as it is more likely to observe distances of $d=1$ than $d=0$ we find that modifications to $w(d=1)$ result in more changed outcomes compared to modifications to $w(d=0)$.
As $k$ increases, this mismatch decreases, approaching a correlation coefficient of $0.9$ at $k=15$. The same percentage is found with the 2022 version of the VAA.
These results illustrate that as the number of recommended political candidates increases, the effective impact of the algorithm modifications decreases. 
Although showing more candidates does not completely solve the robustness issue, it greatly reduces it. Even just showing the top $k=6$ candidates a user most agrees with raises the correlation coefficient to approx. $0.8$ for low distance weights ($w(d=0)$ and $w(d=1)$). Again, this is consistent with the percentage found with the 2022 version of the VAA. 

\subsection{Impact of removing questions}
\begin{figure}[h]
    \centering
     \includegraphics[width=\textwidth]{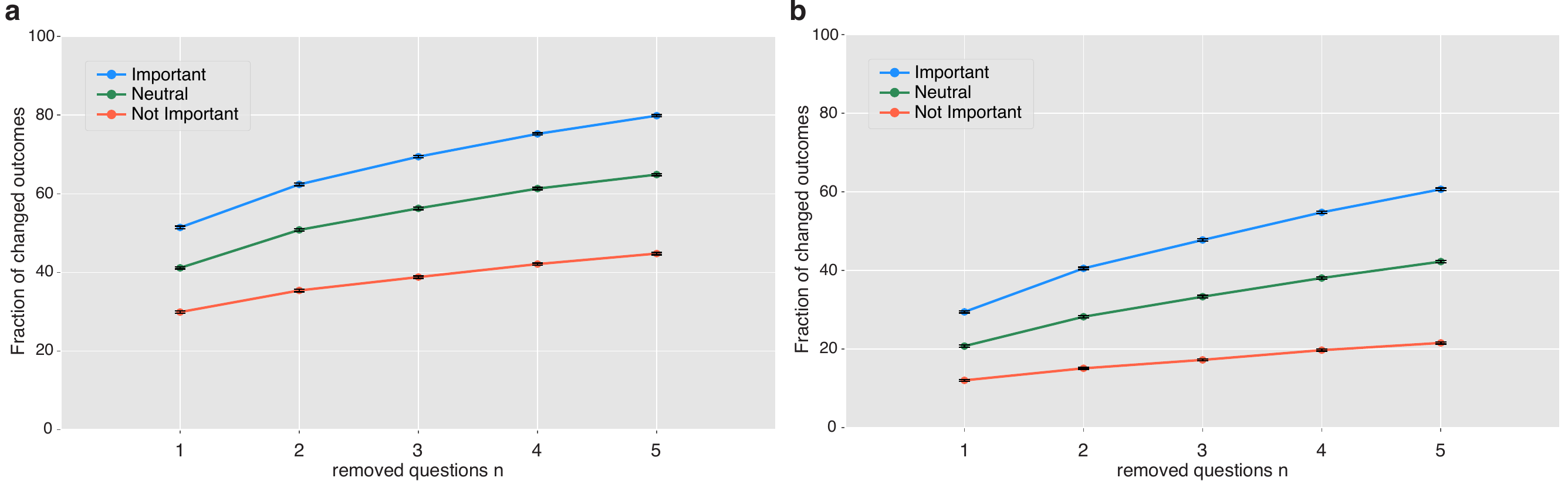}
    \caption{Impact of removing $n$ questions from the agreement calculation. We split the analysis up into different question importance-types. Error bars denote 95\% confidence intervals across different sets of questions. \textbf{a}, Results at candidate level. \textbf{b}, Results at party level.}
    \label{fig:fig6}
\end{figure}

The results from modifying weights might seem drastically high.
To put them into context we compare the effects to dropping questions (Fig.~\ref{fig:fig2}e).
In this case we use the original weights, but calculate agreement percentages with $n$ questions removed. 
Fig.~\ref{fig:fig6} shows the results split up according to question importance.
At candidate level (Fig.~\ref{fig:fig6}a) we find that just removing a single `important' question can drastically change the outcome, with more than 50\% of outcomes being affected.
Dropping three questions changes 70\% of outcomes, and dropping five questions 80\% of outcomes.
The results are less extreme for questions marked as `neutral' and `not important'; however, dropping one question still changes more than 40\% of outcomes for neutral, and approximately 30\% of outcomes for not important questions. 
Compared to modifying single weights with $\delta$, dropping one `not important' question has the same impact as modifying weights $w(d=0)$ or $w(d=1)$ by $\delta = 2$ (Fig.~\ref{fig:fig3}a). 
Dropping one `neutral' or `important' question goes beyond $\delta=3$ edits.

At the party level, the results are less extreme, but still disturbing (Fig.~\ref{fig:fig6}b). Dropping one `important' question changes $\sim$30\% of outcomes, meaning the \textit{Kandidattest} will recommend candidates from a different political party almost a third of the time.
Results are less drastic for `neutral' and `not important' questions, however, they change approximately 20\% and 12\% of outcomes from one dropped question. 
Compared to the results we see from modifying weights, dropping just one question has a more dramatic effect on outcomes. Results for the 2022 \textit{Kandidattest} are qualitatively similar. 

\subsection{Comparing results to `real-world users'}
\begin{figure}[h]
    \centering
     \includegraphics[width=0.8\textwidth]{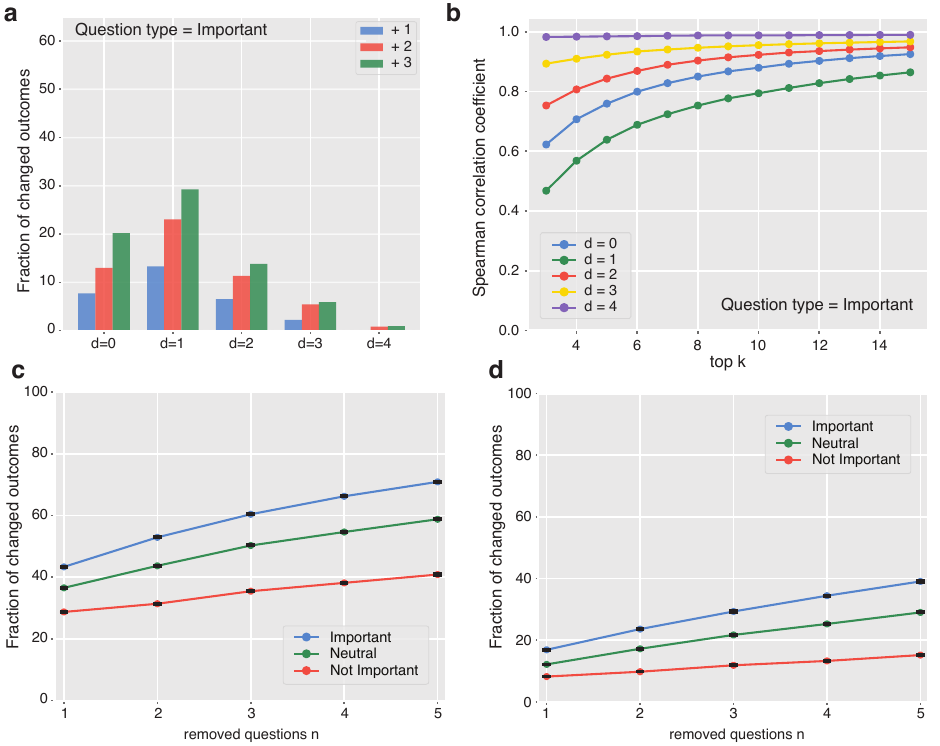}
    \caption{Results for `real-world users'. \textbf{a}, Percentage of changed outcomes as result of single modification. Results are on political party level, and all questions have been marked as `important'. No confidence intervals are computed as we treat all users as a single batch. \textbf{b}, Robustness of top-$k$ ranked candidates. All questions are marked as `important' and results are for a single $\delta = 3$ modification of the weights. \textbf{c}, Impact of removing $n$ question (while using the original weights of the algorithm). Results on candidate level, with errorbars showing 95\% confidence intervals across removal of different sets of questions. \textbf{d}, Impact of removing $n$ question on political party level.}
    \label{fig:fig_real_world_users}
\end{figure}

Until now we have focused on `random users', however, these are likely not capturing the possible correlations between responses to questions given in the real-world. 
For instance, the \textit{Kandidattest} includes multiple questions on related policy issues that real people might answer similarly to. 
To account for this, we create `real-world users' by modifying responses from political candidates (where answers are offset by factors of $-1$ and $+1$).
Since we do not have any information on how these individuals might rate questions, the following analyses are performed with all questions marked with a single importance type.
Fig.~\ref{fig:fig_real_world_users} shows the results for these `real-world' users.

Fig.~\ref{fig:fig_real_world_users}a shows the impact of single modifications for `real-world' users on political party level. 
A $\delta=1$ change to $w(d=0)$ changes approximately 8\% of outcomes on political parties, when all questions are marked as `important', while modifications to $w(d=1)$ change 13\% of outcomes. This is in line with our past explanation that distances of $d=1$ are more likely to occur compared to $d=0$, resulting in changes in $w(d=1)$ having a larger impact.
This is qualitatively similar to what we observed for `random users', but the magnitudes are smaller.

For the top-$k$ analysis we find qualitatively similar results as for `random users', however, differences are more pronounced for `real-world users' (Fig.~\ref{fig:fig_real_world_users}b).
This is due to all questions being marked as important for `real world users'.
For $k = 3$ there are large differences as correlation coefficients for $w(d = 0)$ and $w(d = 1)$ are respectively 0.43 and 0.62. This illustrates a large mismatch between the recommendations of the modified and original algorithms. However, modifying $w(d = 4)$ has negligible impact on the output.
Similar to `random users', as $k$ is increased the correlations approach 0.9.

When dropping $n$ questions we find comparably similar results compared to Fig.~\ref{fig:fig6}, backing up our previous findings that the \textit{Kandidattest} is very unstable when questions are removed. 
Dropping $n=1$ question changes 29\%, 37\% and 43\% outcomes for candidate level respectively for questions rated as `not important', `neutral', and `important' (Fig.~\ref{fig:fig_real_world_users}c).
Evaluating changes on party level, the results are less extreme, however, dropping one `important' question changes 16\% of outcomes (Fig.~\ref{fig:fig_real_world_users}d). 

Taken together, the results for `real-world users' backs up our claims that the \textit{Kandidattest} is an unstable, or brittle, algorithm, where minor changes to its underlying weights, or a removal of just one question, can produce entirely different outcomes.

\section{Quantifying potential real-world implications of the \textit{Kandidattest}'s brittleness}
We focus here on the results for `random users' as they incorporate the nuances around question importance (for `real-world' users, we used the same importance for all questions).
Our results reveal that minor modifications to $\delta =1$ in one of the algorithm's weights can change up to 27\% of candidate recommendations, and roughly 9\% of party recommendations. 
To help put this into context, we quantify how many votes this marginally different algorithm could have potentially affected.
According to the Danish Parliament~\cite{folketing}, there were 4,303,429 eligible voters in the 2026 general elections, and of these, 83.98\% actually voted. The Danish National Election Study surveyed a representative sample of voters in 2022 and found that 62\% of voters used a VAA. Approximately 45\% of respondents who used a VAA also reported following the algorithm's advice ~\cite{DanishNationalElectionStudy}.
This puts the number of voters who could potentially be affected by VAA to be slightly above one million.
To quantify how many of these could have been affected, we focus on our result for single modifications ($\delta=1$) and party level statistics (Fig.~\ref{fig:fig3}b), as these quantify changes in how political parties are recommended.

Assuming a modification could have occurred in any of the algorithm's weights, this puts the range of changed outcomes to lie in the range between 1.38\% ($w(d=4, I =\mathrm{\textit{neutral}})$) and 9.42\% ($w(d=0, I =\mathrm{\textit{important}})$).
Assuming further that people used the \textit{Kandidattest} (or other VAAs that behave similarly) to form an opinion on who to vote for, this puts the number of potentially affected voters, and thereby votes, to be between 14.000 and 95.000. This corresponds to 0.39\% to 2.63\% of cast votes.
Converting votes to the number of seats in the parliament (where one mandate usually requires between 17.000 and 20.000 votes), this puts the effect of the modification conservatively to be between 0 mandates and 4-5 mandates. 

To estimate the effect of dropping one question (Fig.~\ref{fig:fig6}b), we first calculate the average percentage of outcomes that are changed by dropping it.
We do this as users individually select a question as important, neutral, or not important. One person might select a specific question as important, while another user might select it to be neutral.
As such, we average over the three curves in Fig.~\ref{fig:fig6}b (29.48\% for important, 20.78\% for neutral, and 12.07\% for not important), and find that dropping one question, on average, changes 20.77\% of outcomes.
Repeating the calculations from above, this corresponds to roughly 230.000 votes, corresponding to 11-13 mandates.
The government formed after the 2026 elections is a minority coalition, lacking 8 mandates, and relying on votes from other parties to achieve a majority. This potentially means that both a small $\delta=1$ modification in the correct weight, and dropping $n=1$ questions could have affected election results, as it could have affected which parties formed the minority government.

\section{Discussion and conclusion}
Our analysis shows that the \textit{Kandidattest} is highly sensitive to small (and plausible) adjustments of algorithmic weights and to the removal of individual questions. Even minimal changes lead to different candidate or party recommendations in a substantial fraction of cases. From a technical point of view, this shows that the system cannot be considered robust.

When placed in the context of prior work on the effects of VAAs, our findings take on additional significance. Previous studies have shown that VAAs influence vote choice and party switching~\cite{germann2019getting}. In the Danish context, Tromborg and Albertsen \cite{tromborg2023candidates, munzert2021meta} show that VAA users who received advice incongruent with the party they expected to vote for became 16\% more likely to change their vote intention. They estimate that this mechanism could have affected approximately 110,000 votes in the 2022 general election, corresponding to around five parliamentary mandates. Our results are of a comparable order of magnitude. We show that marginal technical changes could plausibly affect tens of thousands of votes, and that dropping a single question could affect around 230,000 votes. Together, these findings suggest that VAAs combine two properties: users are responsive to their recommendations, and the recommendations themselves are sensitive to small technical decisions. This indicates that the impact of VAAs can be huge: in the Danish case, VAAs had the potential to shift the overall electoral results in both the 2026 and 2022 elections. In 2026, both minor modifications to the algorithm's weights, and dropping a question could have impacted government formation; in 2022 the government coalition formed had a majority of only 3 mandates, which is less than the possible swing of mandates occurring with both minor modifications and by dropping a question.

Prior work shows that the effects of VVAs are strongest among undecided voters and voters without strong prior identification~\cite{tromborg2023candidates}. These voters are more likely to accept algorithmic advice and to treat algorithmic outputs as authoritative. Banning VAAs is unlikely to resolve this issue as these users will rely on other digital services for political information. Hence, it becomes particularly important to ensure that VAAs are robust to use. 
For example, the Dutch data protection authority recently warned voters about using AI chatbots for voting advice, citing that they are biased, and predominantly recommend the same two parties, irrespective of the users' question or instruction~\cite{ap}.
A related study also found that 1 in 10 Dutch citizens are likely to ask AI for election advice~\cite{algosoc}.
In addition, there have been attempts to use various machine learning models --- including large language models --- for designing VAA questions, correlating answers to single questions to party affiliation, and predicting users' answers to future questions based on past answers~\cite{buryakov2024enhancing, kleiner}, even if interpreting the outputs of these models is rather hard~\cite{loyola2019black, dobson2023reading}.
It has also been proposed to use conversational agents to help users understand VAA statements~\cite{gemenis2024artificial}. However, LLMs have been proven unreliable in the context of elections. In general, their reliability is harmed by the inevitability of errors, or so-called `hallucinations'~\cite{aiforensics, xu2024hallucination}.
These points illustrate that improving the robustness of VAAs is only one part of the challenge. 
Any algorithmic tools used for voting advice should be classified as high-risk systems under the EU's AI Act~\cite{aiact}, and undergo regular audits to ensure they are unbiased, robust, do not mislead, are safe to use, and do not affect our democracies. 
It is not difficult to imagine a bad actor trying to influence elections through VAAs.
In fact, an individual working for one of the right-wing parties developed his own VAA during the 2026 Danish general elections. The VAA was highly criticized as it overwhelmingly recommended right-wing parties, phrased questions in a misleading way, and did not disclose the political ties of its creator~\cite{dr-df-valgtest}. 

When considering the role of automated systems in our daily lives, several scholars have advocated increasing the public's `algorithmic literacy'~\cite{koenig2020algorithms, cotter2020algorithmic, klawitter2018s}. Evidence shows that lower-literacy users use algorithms as black-box systems whose operations are too technical, complicated, and vague to understand. Furthermore, while people with high-literacy focus on fairness, accountability, transparency, and explainability (FATE), people with low literacy are significantly more concerned with performance values~\cite{shin2022seeing}. This aligns with the common misconception that algorithms are inherently right~\cite{garfinkel2017toward, pan2007google}.
In such settings, the inner workings of each VAA should be properly described in plain language to users so that they, regardless of expertise, can understand their shortcomings and limitations.

As the \textit{Kandidattest} algorithm cannot be considered robust, several practices could be implemented to overcome the system's limitations. First of all, showing the first \textit{k} politicians and not only the top one would provide users with a more well-rounded feedback. This practice would be even more effective in our eyes without showing the percentages of agreement. This could lead users, in fact, to check the answers of the political candidates, helping them to make a more informed decision about who they should vote for. The bottom line is that VAAs should be aiming less to reach a quantitative ranking of candidates, and more to empower an informed choice between them.

Another possible improvement could be to provide users with a list of candidates who think alike, either considering, or not, their closeness to the user. This approach interprets each politician as a point in a space that can be further or closer to where the user is located based on the agreement percentage~\cite{teran2012using}. The space could be segmented in order to output \textit{n} close political candidates, maybe the ones that are closest to the user, or to output different groups of similar \textit{n} candidates. Further work could focus on how the candidate space is built by the current algorithm, and on what are other possible ways of building this space.

Our results on removing questions illustrate that removing one question can drastically alter outcomes of the VAA. The reverse is also true. Adding additional questions to the system can make it more robust. 
There is, of course, a trade-off between ease of use and questionnaire length. However, we believe that adding up to five additional questions will not put an extra undue burden on the user.

Lastly, our work should not be seen as a critique of Politiken as a media organization. Politiken was one of the few media companies that responded to our freedom of information request and was transparent in sharing both source code and documentation. They should be acknowledged for this, as other media outlets we reached out to refused our freedom of information requests, saying that their algorithms were proprietary. Our critique instead pertains to VAAs more broadly. Similar architectural and design choices are common across European VAAs, suggesting that the robustness issues identified here are unlikely to be unique to the \textit{Kandidattest}.
We hope that these results will encourage and facilitate future audits of VAAs across Europe.

\section{Implications and recommendations}
There is a need for VAA systems. In busy day-to-day lives, these systems help voters navigate the many choices in multi-party democracies. Below we present recommendations that we believe can drastically improve the robustness of the \textit{Kandidattest}, and other VAAs similar to it. Our recommendations are:

\begin{itemize}
    \item VAAs need to be constantly monitored, audited, and evaluated. How users interact with these systems should also be continuously studied.
    \item We recommend that at least 25 questions should be used to calculate the agreement between users and political candidates. Adding more questions increases the robustness of VAA systems.
    \item Answers should be collected at the political candidate level, not the political party level. Doing this will reduce the volatility of the VAA system.
    \item The use of quantitative metrics should be reduced and  multiple candidates should be presented to users. In addition, information on how recommended candidates agree and disagree on various policy issues should be provided to the user. This would empower the user to read the candidates' proposed policies and make an informed choice between them.
    \item  Explanations of the inner workings of VAAs should be provided to users, explaining, in plain language, the limitations and pitfalls of the algorithmic system.
    \item  VAAs provide voting advice and their recommendations can have election-altering results. As such, VAAs should meet the strict requirements that apply to high-risk systems under the EU AI Act.
\end{itemize}

\section*{Generative AI Usage Statement}
No generative AI tools have been used to write the paper or analyze data, however, GenAI tools have been used as assistance for code writing and completion. For the latter, they have been used with human oversight, and we have tested and verified that the code works correctly. 

\bibliography{scibib}
\bibliographystyle{Science}

\end{document}